\documentclass[letterpaper, 10pt, conference]{ieeeconf}  

\IEEEoverridecommandlockouts                              

\usepackage{graphicx}
\usepackage{url}  
\usepackage{amsmath}
\usepackage{todonotes}

\begin{document}

\title{\LARGE \bf
	Low-complexity deep learning frameworks for acoustic scene classification using teacher-student scheme and multiple spectrograms 
}

\author{Technical Report for DCASE 2023 Task 1 challenge (AIT-Essex-VNU-System)\\
Lam Pham, Dat Ngo, Cam Le, Anahid Jalali, Alexander Schindler
	\thanks{L. Pham, A. Jalali, and A. Schindler are with Austrian Institute of Technology (AIT), Austria. D. Ngo is with University of Essex, UK. C. Le is with Ho Chi Minh University of Technology, Vietnam.}%
}

\maketitle
\thispagestyle{empty}
\pagestyle{empty}

\begin{abstract}
In this technical report, a low-complexity deep learning system for acoustic scene classification (ASC) is presented. 
The proposed system comprises two main phases: (Phase I) Training a teacher network; and (Phase II) training a student network using distilled knowledge from the teacher.
In the first phase, the teacher, which presents a large footprint model, is trained.
After training the teacher, the embeddings, which are the feature map of the second last layer of the teacher, are extracted. 
In the second phase, the student network, which presents a low complexity model, is trained with the embeddings extracted from the teacher.
Our experiments conducted on DCASE 2023 Task 1 Development dataset have fulfilled the requirement of low-complexity and achieved the best classification accuracy of 57.4\%, improving DCASE baseline by 14.5\%.

\indent \textit{Clinical relevance}--- Mixup augmentation, Convolutional Neural Network (CNN), spectrogram, late fusion.

\end{abstract}

\section{Introduction}
\label{intro}

To deal with the ASC challenge of mismatched recording devices, the state-of-the-art systems mainly leverage ensemble techniques: Ensemble of spectrogram inputs~\cite{lam01, lam03, phan2019spatio, lam04, lam05, truc_dca_18, yuma, huy_mul, lam02, pham2022wider, pham2021deep, pham2021dcase} or ensemble of different classification models~\cite{phaye_dca_18, zhao_dca_17, hong_dca_18}. 
Although these approaches prove effective to deal with the issue of mismatched recording devices and achieve potential results, they present large model complexity.
This lead to challenges to apply ASC components on edge-devices.
Recently, DCASE 2021 Task 1A challenge~\cite{task1a_2022} focuses on dealing the issue of high-complexity model.
The challenge requires the maximum model complexity of 128 KB.
Furthermore, the next challenges of DCASE 2022 Task 1 and DCASE 2023 Task 1 do not allow to use pruning techniques as the pruning parameters still occupy the memory and cost the computation on edge-devices.
These challenges also require the maximum MACs (Multiply-Add cumulation) of 30 M.

In this technical report, a low-complexity deep learning frameworks using teacher-student scheme and multiple spectrograms for ASC task is presented.

\section{The teacher network architecture}
\begin{figure}[t]
	\centering
    \scalebox{1.0}{
	\centerline{\includegraphics[width=\linewidth]{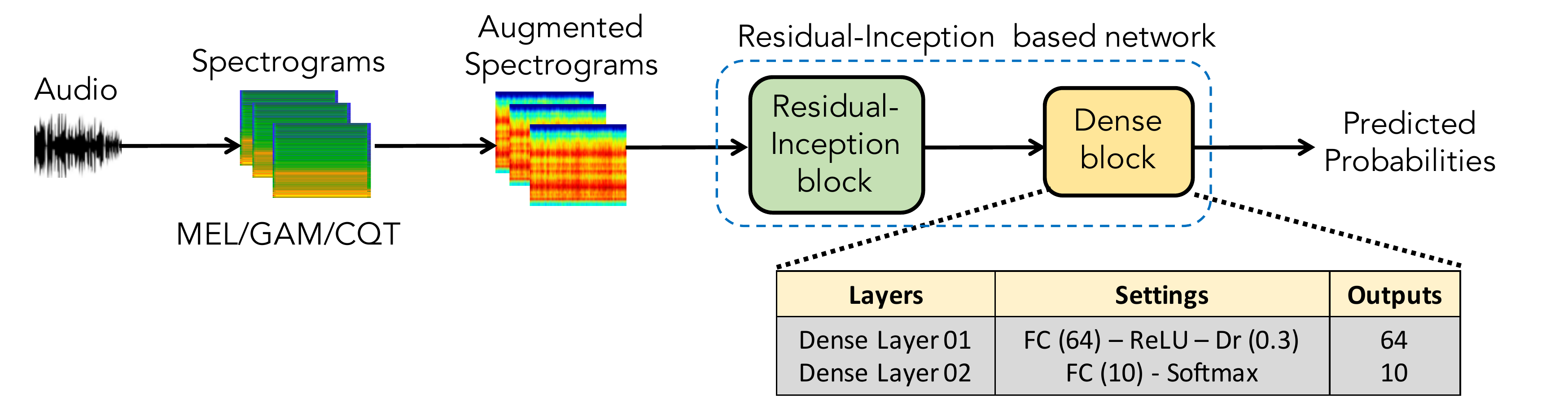}}
	}
	\caption{The high-complexity teacher network architecture.}
	\label{fig:f2}
\end{figure}
\begin{table}[t]
    \caption{The low-complexity student network architecture.} 
        	\vspace{-0.2cm}
    \centering
    \scalebox{0.7}{
    \begin{tabular}{|l |c|} 
        \hline 
            \textbf{Layers}   &  \textbf{Output}  \\
        \hline 
        Input & $128{\times}128{\times}3$ \\
         Convolution ([$2{\times}2$] $@$  C\_{out}1=16) -  ReLU - BN - AP [$2{\times}2$]- Dropout (10\%) & $64{\times}64{\times}16$\\
         
         Convolution ([$2{\times}2$] $@$  C\_{out}2=16) - ReLU - BN - AP [$2{\times}2$] - Dropout (15\%) & $32{\times}32{\times}16$\\
         
         Convolution ([$2{\times}2$] $@$ C\_{out}3=16) - ReLU - BN - AP [$2{\times}2$] -Dropout (20\%) & $16{\times}16{\times}32$\\
         
         Convolution ([$2{\times}2$] $@$ C\_{out}4=32) - ReLU - BN - GAP - Dropout (25\%) & $32$\\
                          
         FC (64) - ReLU - Dropout (30\%)  &  $64$       \\
         FC (10) - Softmax  &  $10$       \\

       \hline 
    \end{tabular}
    }
    \label{table:M2} 
\end{table}

As Fig.~\ref{fig:f2} shows, the proposed teacher network can be separated into three main steps: the front-end feature extraction, the online data augmentation, and the convolutional neural network (CNN) based network.
Initially, a raw audio signal is firstly transformed into three spectrograms of  128$\times$132 by using MEL filter~\cite{librosa_tool}, Gammatone filter~\cite{aud_tool}, or CQT~\cite{librosa_tool} with the FFT number, Hanna window size, hop size, and the filter number set to 4096, 2048, 326, and 128. 
Next, we apply delta  and delta-delta on these spectrograms to generate three-dimensional spectrograms of 128$\times$128$\times$3.
The original spectrogram, delta, and delta-delta).
We then apply the Mixup~\cite{mixup1, mixup2} augmentation method on the spectrograms.
We finally feed the augmented spectrograms into back-end deep learning networks for classification, referred to as the teachers. 
As we use three spectrogram input, we train three individual teachers.

\begin{figure}[t]
	\centering
    \scalebox{1.0}{
	\centerline{\includegraphics[width=\linewidth]{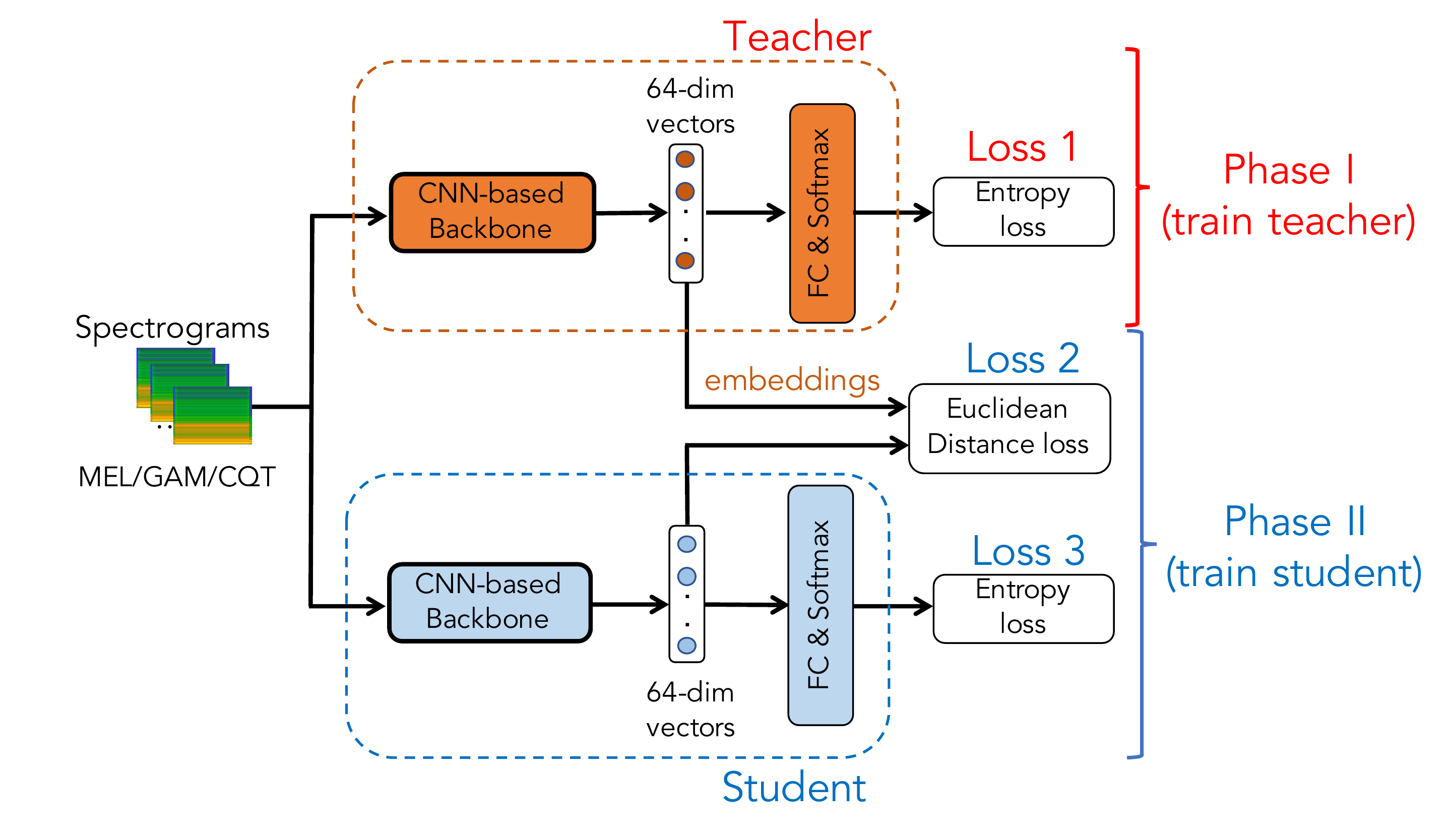}}
	}
	\caption{Training the student network using knowledge distillation.}
	\label{fig:f1}
\end{figure}

Regarding the teacher architecture, it comprises two main parts: a CNN-based backbone followed by a dense block.
The CNN-based backbone, which presents a residual-inception based architecture, is reused from~\cite{pham_j2, pham2022wider}.
The dense block comprises two dense layers (Dense Layer 01 and Dense Layer 02), which is shown in the lower part of Fig.~\ref{fig:f1}.
After training the teachers, the embeddings, which are the feature map at the first fully connected layer of the dense block (FC (64)), are extracted for training the student networks.
The teachers are trained using Entropy loss ($Loss_1$) as shown in Fig.~\ref{fig:f1}.
\begin{table*}[t]
    \caption{Performance (Acc.\%/Log loss) comparison among DCASE baseline, the individual students without distillation, ensemble of three students without distillation, the individual students with distillation, the ensemble of three students with distillation} 
        	\vspace{-0.2cm}
    \centering
    \scalebox{0.9}{
    \begin{tabular}{| l |  c |  c c c  c |   c  c  c  c| } 
        \hline 
                         &\textbf{DCASE}     &\textbf{CQT}      &\textbf{GAM}      &\textbf{MEL} & \textbf{Ens. Students}   &\textbf{CQT}      &\textbf{GAM}      &\textbf{MEL} & \textbf{Ens. Students} \\
                                     &\textbf{baseline}  &\textbf{w/o dis.} &\textbf{w/o dis.} &\textbf{w/o dis.}  &\textbf{w/o dis.}  &\textbf{w/ dis.} &\textbf{w/ dis.} &\textbf{w/ dis.}  &\textbf{w/ dis.} \\
        \hline 
        \hline

        Airport           &39.4/1.534           &46.6/2.091 &49.1/1.694 &45.7/1.801   &57.1/1.327 &52.6/2.112 &38.2/1.789 &51.7/1.656 &62.3/1.306      \\
        Bus               &29.3/1.758           &63.0/1.793 &61.8/1.559 &68.6/1.554   &75.4/0.880 &67.7/1.700 &68.0/1.583 &55.9/1.612 &77.5/0.841        \\
        Metro             &47.9/1.382           &24.9/2.130 &33.7/1.745 &42.1/1.668   &43.8/1.363 &36.6/2.095 &33.4/1.801 &37.1/1.708 &45.8/1.305      \\
        Metro station     &36.0/1.672           &20.7/2.113 &35.3/1.853 &37.0/1.848   &44.4/1.510 &25.0/2.109 &41.8/1.930 &51.4/1.755 &47.0/1.453     \\
        Park              &58.9/1.448           &56.9/2.037 &67.6/1.716 &76.4/1.554   &76.6/1.118 &66.6/2.052 &74.3/1.650 &73.1/1.478 &78.9/1.037      \\
        Public square     &20.8/2.265           &16.5/2.123 &37.1/1.915 &37.8/1.890   &40.9/1.590 &12.4/2.163 &32.4/1.950 &31.9/1.869 &39.9/1.646       \\
        Shopping mall     &51.4/1.385           &38.0/2.108 &73.8/1.686 &61.3/1.747   &69.6/1.209 &23.1/2.131 &75.5/1.741 &60.6/1.619 &63.7/1.278      \\
        Street pedestrian &30.1/1.822           &19.4/2.279 &23.4/1.927 &15.2/2.026   &24.2/2.140 &15.9/2.219 &27.6/2.004 &27.9/1.911 &23.4/2.134      \\
        Street traffic    &70.6/1.025           &56.6/2.082 &73.1/1.519 &75.5/1.425   &77.7/0.936 &50.1/2.108 &77.1/1.458 &73.5/1.552 &77.6/0.948      \\
        Tram              &44.6/1.462           &34.7/2.083 &29.0/1.688 &43.1/1.651   &58.0/1.413 &36.6/2.006 &50.9/1.749 &45.3/1.692 &58.2/1.384     \\
        \hline  
        \hline                                                    
        Average          &42.9/1.575            &37.7/2.084 &50.7/1.730 &50.2/1.716  &56.8/1.349  &38.6/2.070  &51.9/1.765 &50.8/1.685 &57.4/1.333        \\
        \hline         
        Memory (KB)      &46.5                  &28.8       &28.8       &28.8        &88.7         &28.8       &28.8       &28.8        &88.7     \\
        \hline        
        MACs (M)         &29.23                 &9.75       &9.75       &9.75        &29.27        &9.75       &9.75       &9.75        &29.27   \\
        \hline                                                   

    \end{tabular}
    }
    \label{table:res01} 
\end{table*}

\section{The student network architecture}
A student network architecture is presented in Table~\ref{table:M2}.
As we use three spectrograms, we develop three individual students which share the same network architecture.
As the configuration shows in Table~\ref{table:M2}, three student presents 22962 trainable parameters, which occupy 88704 Byte (one parameter is presented by 32 bit) and 29267550 MACs on an edge device.
Training the students is presented in Fig.~\ref{fig:f1} with two loss functions ($Loss_2$ and $Loss_3$).
The $Loss_3$ is traditional Entropy loss which is applied to the final layer (Softmax layer) for classification.
Meanwhile, the $Loss_2$ is mean squared error (MSE) which is applied to the fully connected layer (FC (64)) of the student and the 64-dimensional embeddings extracted from the teacher.
During training the student, the Mixup data augmentation is not applied and the ratio of $Loss_2$ and $Loss_3$ is empirically set to 1:1.

As we apply three spectrograms of CQT, GAM, and Mel, we fuse the probability results obtained from three individual students. 
In particular, we conduct experiments over individual student network with each spectrogram input, then obtain predicted probability of each network as  \(\mathbf{\bar{p_{s}}}= (\bar{p}_{s1}, \bar{p}_{s2}, ..., \bar{p}_{sC})\), where $C$ is the category number and the \(s^{th}\) out of \(S\) networks evaluated. 
Next, the predicted probability after PROD fusion \(\mathbf{p_{f-prod}} = (\bar{p}_{1}, \bar{p}_{2}, ..., \bar{p}_{C}) \) is obtained by:

\begin{equation}
\label{eq:mix_up_x1}
\bar{p_{c}} = \frac{1}{S} \prod_{s=1}^{S} \bar{p}_{sc} ~~~  for  ~~ 1 \leq s \leq S 
\end{equation}

Finally, the predicted label  \(\hat{y}\) is determined by \begin{equation}
    \label{eq:label_determine}
    \hat{y} = arg max (\bar{p}_{1}, \bar{p}_{2}, ...,\bar{p}_{C} )
\end{equation}

\section{Evaluation Setting and Results}

\subsection{TAU Urban Acoustic Scenes 2022 Mobile, development dataset~\cite{dc_2021_1A}}
This report presents the results on DCASE 2023 Task 1 Development set, which was proposed in DCASE 2023 challenge~\cite{dcase_web}. 
In this challenge, the limitation of model size is set to 128 KB of trainable parameters and the maximum MACs is set to 30 M, not allow to use pruning techniques, and evaluate on 1-second audio segment.
The dataset is slightly unbalanced, being recorded across 12 large European cities: Amsterdam, Barcelona, Helsinki, Lisbon, London, Lyon, Madrid, Milan, Prague, Paris, Stockholm, and Vienna. 
It consists of 10 scene classes: airport, shopping mall (indoor), metro station (underground), pedestrian street, public square, street (traffic), traveling by tram, bus and metro (underground), and urban park.
The audio recordings were recorded from 3 different physical devices namely A (10215 recordings), B (749 recordings), C (748 recordings).
Additionally, synthetic data for mobile devices was created based on the original recordings, referred to as S1 (750 recordings), S2 (750 recordings), S3 (750 recordings), S4 (750 recordings), S5 (750 recordings), and S6 (750 recordings). 
  
To evaluate, we follow the DCASE 2023 Task 1 challenge~\cite{dcase_web}, use two subsets known as Training (Train.) and Evaluation (Eval.) from the Development set for training and testing processes, respectively.
Notably, two of 12 cities and S4, S5, S6 audio recordings are only presented in the Eval. subset for evaluating the issue of mismatched recording devices and unseen samples.

\subsection{Network Implementation}
All the CNN architectures are conducted by Tensorflow frameworks. 
Training CNN architecture use Adam algorithm for the optimization.
We run all experiments on the GPU GeForce RTX 2080.

\subsection{Experimental Results}
The experimental results are presented in Table~\ref{table:res01}.
As Table~\ref{table:res01} shows, results on GAM and MEL are competitive and outperform the records of DCASE baseline and CQT spectrogram.
The ensemble of three models without and with using knowledge distillation achieve accuracy of 56.8\% and 57.4\%, respectively.
The best model using ensemble of multiple spectrogram and knowledge distillation improves the DCASE baseline by 14.5\% 
The log-loss score of this system presents 1.333 which is less than the DCASE score of 1.575.
However, this system requires more memory of 88.7 MB compared with DCASE baseline of 46.5 MB.

\section{Conclusion}

We have presented a low-complexity system for ASC task, which leverages teacher-student scheme and multiple spectrogram inputs.
Our proposed low-complexity ASC system achieves an accuracy of 57.4\%, a log-loss score of 1.333, 88.7 KB memory occupation, and 29.27 M MACs.

\bibliographystyle{IEEEbib}
\bibliography{refs}

\begin{thebibliography}{10}

\bibitem{lam01}
L.~Pham, I.~Mcloughlin, Huy Phan, R.~Palaniappan, and A.~Mertins,
\newblock ``Deep feature embedding and hierarchical classification for audio
  scene classification,''
\newblock in {\em International Joint Conference on Neural Networks (IJCNN)},
  2020, pp. 1--7.

\bibitem{lam03}
L.~Pham, I.~McLoughlin, H.~Phan, R.~Palaniappan, and Y.~Lang,
\newblock ``Bag-of-features models based on {C-DNN} network for acoustic scene
  classification,''
\newblock in {\em Proc. AES}, 2019.

\bibitem{phan2019spatio}
Huy Phan, Oliver~Y Ch{\'e}n, Lam Pham, Philipp Koch, Maarten De~Vos, Ian
  McLoughlin, and Alfred Mertins,
\newblock ``Spatio-temporal attention pooling for audio scene classification,''
\newblock in {\em Proc. INTERSPEECH}, 2019, pp. 3845--3849.

\bibitem{lam04}
Lam Pham, Ian Mcloughlin, Huy Phan, and Ramaswamy Palaniappan,
\newblock ``A robust framework for acoustic scene classification,''
\newblock in {\em Proc. INTERSPEECH}, 2019, pp. 3634--3638.

\bibitem{lam05}
D.~Ngo, Hao Hoang, A.~Nguyen, Tien Ly, and L.~Pham,
\newblock ``Sound context classification basing on join learning model and
  multi-spectrogram features,''
\newblock {\em ArXiv}, vol. abs/2005.12779, 2020.

\bibitem{truc_dca_18}
Truc Nguyen and Franz Pernkopf,
\newblock ``Acoustic scene classification using a convolutional neural network
  ensemble and nearest neighbor filters,''
\newblock in {\em Proc. DCASE}, 2018, pp. 34--38.

\bibitem{yuma}
Yuma Sakashita and Masaki Aono,
\newblock ``Acoustic scene classification by ensemble of spectrograms based on
  adaptive temporal divisions,''
\newblock Tech. {R}ep., DCASE Challenge, 2018.

\bibitem{huy_mul}
Huy Phan, Huy Le~Nguyen, Oliver~Y. Chén, Lam Pham, Philipp Koch, Ian
  McLoughlin, and Alfred Mertins,
\newblock ``Multi-view audio and music classification,''
\newblock in {\em ICASSP 2021 - 2021 IEEE International Conference on
  Acoustics, Speech and Signal Processing (ICASSP)}, 2021, pp. 611--615.

\bibitem{lam02}
L.~Pham, Huy Phan, T.~Nguyen, R.~Palaniappan, A.~Mertins, and I.~Mcloughlin,
\newblock ``Robust acoustic scene classification using a multi-spectrogram
  encoder-decoder framework,''
\newblock {\em Digital Signal Processing}, vol. 110, pp. 102943, 2021.

\bibitem{pham2022wider}
Lam Pham, Khoa Tran, Dat Ngo, Hieu Tang, Son Phan, and Alexander Schindler,
\newblock ``Wider or deeper neural network architecture for acoustic scene
  classification with mismatched recording devices,''
\newblock in {\em Proceedings of the 4th ACM International Conference on
  Multimedia in Asia}, 2022.

\bibitem{pham2021deep}
Lam Pham, Alexander Schindler, Mina Sch{\"u}tz, Jasmin Lampert, Sven Schlarb,
  and Ross King,
\newblock ``Deep learning frameworks applied for audio-visual scene
  classification,''
\newblock {\em arXiv preprint arXiv:2106.06840}, 2021.

\bibitem{pham2021dcase}
Lam Pham, Alexander Schindler, Hieu Tang, and Truong Hoang,
\newblock ``Dcase 2021 task 1a: Technique report,''
\newblock {\em Tech. Rep., DCASE2021 Challenge}, 2021.

\bibitem{phaye_dca_18}
Sai Phaye, Emmanouil Benetos, and Ye~Wang,
\newblock ``{SubSpectralNet} using sub-spectrogram based convolutional neural
  networks for acoustic scene classification,''
\newblock in {\em Proc. ICASSP}, 2019, pp. 825--829.

\bibitem{zhao_dca_17}
Zhao Ren, Kun Qian, Yebin Wang, Zixing Zhang, Vedhas Pandit, Alice Baird, and
  Bjorn Schuller,
\newblock ``Deep scalogram representations for acoustic scene classification,''
\newblock {\em IEEE/CAA Journal of Automatica Sinica}, vol. 5, no. 3, pp.
  662--669, 2018.

\bibitem{hong_dca_18}
Hongwei Song, Jiqing Han, Shiwen Deng, and Zhihao Du,
\newblock ``Acoustic scene classification by implicitly identifying distinct
  sound events,''
\newblock in {\em Proc. INTERSPEECH}, 2019, pp. 3860--3864.

\bibitem{task1a_2022}
{Dase Community},
\newblock ``{DCASE 2022 Task 1A Description},''
  \url{https://dcase.community/challenge2022/task-low-complexity-acoustic-scene-classification#description}.

\bibitem{librosa_tool}
Brian McFee, Raffel Colin, Liang Dawen, D.P.W. Ellis, McVicar Matt, Battenberg
  Eric, and Nieto Oriol,
\newblock ``librosa: Audio and music signal analysis in python,''
\newblock in {\em Proceedings of The 14th Python in Science Conference}, 2015,
  pp. 18--25.

\bibitem{aud_tool}
D~P W~(2009) Ellis,
\newblock ``Gammatone-like spectrogram,'' 2009.

\bibitem{mixup1}
Kele Xu, Dawei Feng, Haibo Mi, Boqing Zhu, Dezhi Wang, Lilun Zhang, Hengxing
  Cai, and Shuwen Liu,
\newblock ``Mixup-based acoustic scene classification using multi-channel
  convolutional neural network,''
\newblock in {\em Pacific Rim Conference on Multimedia}, 2018, pp. 14--23.

\bibitem{mixup2}
Yuji Tokozume, Yoshitaka Ushiku, and Tatsuya Harada,
\newblock ``Learning from between-class examples for deep sound recognition,''
\newblock in {\em ICLR}, 2018.

\bibitem{pham_j2}
Lam Pham, Dusan Salovic, Anahid Jalali, Alexander Schindler, Khoa Tran, Canh
  Vu, and Phu~X Nguyen,
\newblock ``Robust, general, and low complexity acoustic scene classification
  systems and an effective visualization for presenting a sound scene
  context,''
\newblock {\em arXiv preprint arXiv:2210.08610}, 2022.

\bibitem{dc_2021_1A}
Annamaria Mesaros, Toni Heittola, and Tuomas Virtanen,
\newblock ``A multi-device dataset for urban acoustic scene classification,''
\newblock in {\em Proceedings of the Detection and Classification of Acoustic
  Scenes and Events (DCASE)}, 2018, pp. 9--13.

\bibitem{dcase_web}
{Detection and Classification of Acoustic Scenes and Events Community},
\newblock {\em {DCASE 2022 challenges}},
\newblock \url {http://dcase.community/challenge2022}.

\end{thebibliography}


\end{document}